# On the complete interface development of Al/Cu magnetic pulse welding via experimental characterizations and multiphysics numerical simulations

J.S. Li[1], T. Sapanathan[2, *], R.N. Raoelison[3], Y.L. Hou[4, **], A. Simar[2], M. Rachik[1]

[1] Université de Technologie de Compiègne, CNRS, Roberval (Mechanics, energy and electricity), Centre de recherche Royallieu - CS 60319 - 60203 Compiègne cedex, France

[2] UCLouvain, Institute of Mechanics, Materials and Civil Engineering, IMAP, 1348 Louvain-la-Neuve, Belgium

[3] Université de Bourgogne Franche-Comté - UTBM, Laboratoire Interdisciplinaire Carnot de Bourgogne, UMR 6303 CNRS, 90100 Belfort, France

[4] School of Mechanical and Power Engineering, Zhengzhou University, Science Road 100, 450001 Zhengzhou, China

Corresponding author: * Thaneshan Sapanathan (thaneshan.sapanathan@uclouvain.be)
** Yuliang Hou (yulianghou@zzu.edu.cn)

**Abstract**

A complex Al/Cu magnetic pulse welding interface is systematically investigated using experimental characterizations and numerical simulations. A Coupled electromagnetic-mechanical simulation is proposed to compute the impact velocity and impact angle along the entire interface. This model allows to further understand the formation mechanism of various interface characteristics during MPW. The results revealed that the impact velocity gradually decreases in conjunction with the gradual increase of the impact angle. These simulations elucidate the experimentally observed successive interface morphologies, i.e., the unwelded zone, vortex zone, intermediate (IM) layers and wavy interface. Microstructural characterizations show that the IM layers are formed by mechanical mixing combined with melting and are characterized by highly heterogeneous porous zone with random sizes and distributions of void. Subsequently, an Eulerian simulation is proposed to investigate the thermomechanical effects during the wave formation which vastly influence the joint quality. The predicted wavy morphologies, temperature distribution and average equivalent plastic strain along the interface provided a better understanding on the formation steps of the experimentally observed wave morphologies. The actual kinematics during wave generation revealed that the wave was formed with repeated deformation of the interface material. The wave amplitude increases with increasing jetting angle; while the wavelength increases with increasing collision point velocity. The ratio between collision point velocity and impact velocity is found to be the most suitable parameter for explaining the influence of the collision parameters on the wavelength. These results indicate that the combination of the Coupled electromagnetic-mechanical simulation and Eulerian simulation can help us investigate the governing mechanisms of complex interface morphologies and further used to optimize the processing parameters during high speed impact welding.

# 1. Introduction

The joints of dissimilar metals are typical examples of hybrid structures, which have growing demand in modern engineering applications as they can offer complementary properties compared to the individual constituent materials. According to Acarer (2012), among the multiple dissimilar metal combinations, aluminum to copper welds are particularly attractive for electric power, electronic and offshore piping applications due to their excellent corrosion resistance, thermal properties and electrical conductivity. However, conventional fusion welding processes can easily introduce thick and brittle Al-Cu intermetallic compound (IMC) layers in those joints. These IMC layers deteriorate the mechanical properties of the joint. Solid-state welding methods are attractive alternatives to conventional processes and offer efficient multi-metallic joints with significant reduction of IMC formation. However,

Avettand-Fènoël and Simar (2016) reported that the common friction-based solid-state welding methods inevitably introduce heat affected zones (HAZ). Impact welding offers an opportunity to obtain high quality dissimilar joints without HAZ if appropriate welding conditions are selected. Zhang *et al.* (2011) reported that impact welding is suitable for welding materials at small and larger scales (millimeters to meters).

Magnetic pulse welding (MPW), one type of impact welding, patented by Lysenko *et al.* (1970), uses pressure generated by an electromagnetic impulse to weld materials. Psyk *et al.* (2011) and Kapil and Sharma (2015) reviewed the process and reported that the MPW applications within the automobile and electrical industries are increasing, owing to its environmental friendliness and flexibility. MPW has been proven effective for tubular joints of carbon fiber-reinforced plastic/aluminum and copper/steel as demonstrated by Cui *et al.* (2019) and Faes *et al.* (2019), sheet joints of aluminum/magnesium and aluminum/copper as presented by Kakizaki *et al.* (2011) and Kwee *et al.* (2016), which contain at least one joining part with good electrical conductivity. Groche *et al.* (2017) have evidenced that, during MPW, impact velocity and impact angle are crucial parameters as they affect the interface microstructure and weld quality. To understand the physics behind MPW, Photon Doppler Velocimetry (PDV) or high-speed camera measurements are used to obtain the impact velocity. By this technique, Lueg-Althoff *et al.* (2018) measured the impact velocity and further used it to analyse the radial flyer kinetics. They claimed that welding can be achieved at low impact velocity with a low frequency pulse generator. Groche *et al.* (2017) used a high-speed camera to record the collision process of aluminum assemblies with a special mechanical test rig. They could thus identify the weldability window based on impact velocity and impact angle. However, these experimental measuring techniques are unable to capture the complete impact velocity field and impact angle along the interface for tubular assemblies with a fieldshaper.

During MPW, the interface zones experience severe plastic deformation producing complex and heterogeneous interfaces. For example, Raoelison *et al.* (2016) found that additional interfacial features like waves, wakes and vortices were produced due to the increase of interfacial instability. Meanwhile, a high intensity of impact enables to produce large amounts of heat at the interface during welding. Li *et al.* (2019) used the kinetics of precipitates and dispersoids in 6000 series aluminum alloy to extract the local temperature at the interface. They found the interface temperature may be larger than 500 °C. Nevertheless, there is generally insufficient time for the heat dissipation before thermal softening which probably results in heat accumulation at the interface. This heat accumulation leads to partial melting and formation of continuous or discontinuous intermediate (IM) layers at the interface as presented by Li *et al.* (2020a). According to Sapanathan *et al.* (2015), these IM layers could critically affect the quality of the welds due to their brittle behavior, the presence of stress concentrations and stress discontinuities.

In MPW, wavy interfaces are resulting in good bonding regardless of the material combination. Raoelison *et al.* (2013) claimed that the tubular MPW joints of aluminum alloys with a wavy interface without voids show a good and permanent bonding. Hahn *et al.* (2016) performed a peel test on the MPW joints and showed that the wavy interface pattern promotes joint strength. Thus, numerous research studies were carried out to better understand the wave formation mechanism in terms of morphology, microstructure and formation kinematics during impact welding. Li *et al.* (2020b) reported that the wavy interface is formed thanks to the presence of an interdiffusion zone either at solid-state or with the emergence of local melting. These findings were supported by transmission electron microscopy (TEM) observations. Cui *et al.* (2014) used an analytical model to predict the wave formation and revealed that a wave with peak height equal to the width of a transition zone is generated by the shear instability at an angle of 4°. Recently, numerical simulations explain the wave formation kinematics. Such data may not be obtained by in-situ experimental methods due to the transient character of the high impact process. Bataev *et al.* (2019) pointed out that the wave formation is due to the sequential indentation of protrusions formed on opposite sides of the welding workpieces. However, Lee *et al.* (2019) stated that the wave morphology results from internal stresses during collision. Effects on wavy morphologies (i.e., wavelength and amplitude) with respect to the collision parameters have also been reported in previous work. Wang *et al.* (2020) claimed that the wavelength increases with decreasing collision velocity and the amplitude increases with increasing jetting velocity. However, the analytical formula combined with the experimental work of Watanabe and Kumai (2009) contradict that wavelength increases with increasing collision velocity. Moreover, Lee *et al.* (2019) suggested that the impact velocity should dictate the wave amplitude.

In Raoelison *et al.* (2015), the effects of welding parameters on the interface behavior for an Al/Cu MPW were investigated via experimental observations, while the governing mechanisms and the associated thermomechanical kinetics were not clarified. These mechanisms are responsible for the formation of interfacial features. Moreover, the formation mechanisms of various types of waves and cavities (i.e. submicron and microscale pores) are not investigated in details in their study. In another work of Raoelison *et al.* (2016), an Eulerian simulation was used to predict a regular wave formation corresponding to an Al/Al interface. However, in the MPW of dissimilar materials, the interface behaviors are more complex due to the mismatch of strain hardening and thermal expansion coefficients, and thus the interface morphologies are not easy to predict. Moreover, in their work, a constant velocity was assumed in the model without considering the local impact conditions.

From the literature mentioned above, although many research studies have been devoted to MPW, some fundamental aspects of the wave formation steps for dissimilar interface need further investigation. In this work, two different numerical models were used along with experimental validations and investigations of the welding tests. A flowchart given in Fig. 1 explains the modelling and validation steps carried out in this study. A coupled electromagnetic mechanical (CEMM) model is proposed to determine the impact angle and impact velocity along the interface during the MPW process. Based on the CEMM model, various welding zones were identified, and have been later identified as responsible for various interface characteristics from the experimental observations. Subsequently, the obtained local impact parameters were introduced into an Eulerian (thermomechanical) model to predict various types of waves to investigate their formation mechanism at the Al/Cu weld. Thus, it provided a better understanding of the impact conditions which are responsible to produce various types of waves and sound welds.

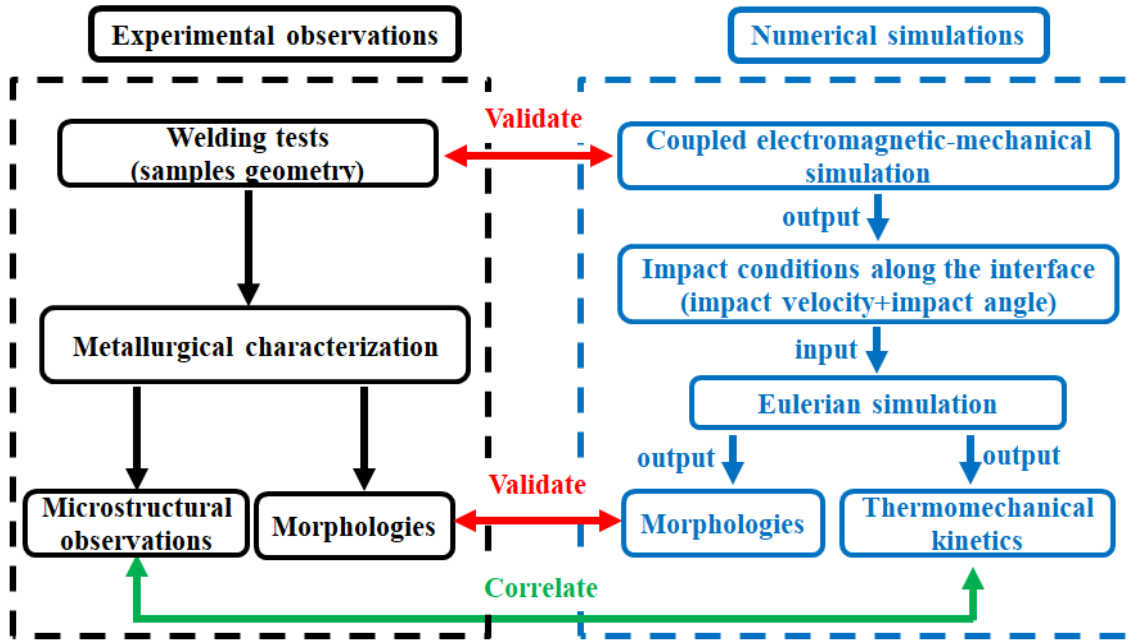

Fig. 1. Flowchart showing various experimental and numerical steps in this study.

## 2. Experimental methods

### 2.1 Welding conditions

Tubular assemblies of Al6060-T6 flyer and copper rod were made by MPW. The welding process was conducted on a PULSAR® system equipped with 690 μF capacitor bank that generates 22 kHz pulse current. A 3D schematic illustration of the MPW setup containing a one turn coil and a CuBe2 fieldshaper is illustrated in Fig. 2a. The workpieces were placed with an overlap configuration of 10 mm as illustrated in Fig. 2b. Welding was performed with an input voltage and initial air gap of 6 kV and 1.64 mm, respectively.

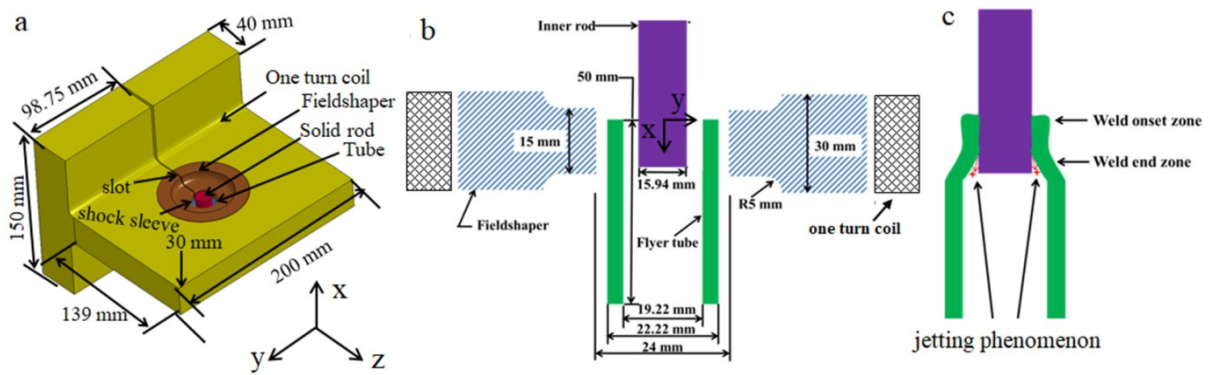

Fig. 2 Schematic of MPW process: (a) 3D isometric projection view of the MPW platform, cross-sectional views of welding configuration (b) before welding and (c) during welding propagation.

## 2.2 Metallurgical characterization

The welded specimen was dissected by a micro chain saw cutting machine to prepare samples for microstructural characterization followed by standard metallurgical sample preparation techniques. Interface morphology and microstructural observations were made using ZEISS Ultra 55 Scanning electron microscope (SEM) with both secondary electrons (SE) and backscattered electrons (BSE) modes.

## 3. Numerical simulations

## 3.1 Coupled electromagnetic-mechanical simulation

A coupled electromagnetic-mechanical (CEMM) model was developed within LS-DYNA® package to compute the spatial distribution of the vertical component of the impact velocity ($V_y$) and impact angle ($α$) along the interface during the welding process. The CEMM model was validated based on the final geometry under various welding conditions (see for example Fig. S1 in the Supplementary Material). The boundary element method (BEM) and finite element method (FEM) were used in the computation of CEMM model. BEM was used to compute the surface current and electromagnetic field, and FEM was used to compute the eddy current and Lorentz force in the workpieces. The CEMM model uses a full-scale 3D geometry (Fig. 2a). It contains 206910 eight-node solid elements. The boundary conditions on the fieldshaper and the workpieces are shown in Fig. 3a, where one corner point 'P' was fixed in all directions and the surface marked with ⊗ was fixed in the x direction (direction normal to the plane) for the fieldshaper. A thin polymer shock sleeve (see Fig, 2a and Fig. 3a) was added in the model between the fieldshaper and the coil, which can effectively avoid electrical contact and maintain the concentricity of the fieldshaper. The top side of the rod and the bottom side of the tube were completely fixed in all directions. The input current used in the CEMM model is given in Fig. 3b. More details about the governing equations, input mechanical and electromagnetic parameters for each part used in the CEMM model are provided in the Supplementary material (Table S1).

Once the accelerating flyer impacts the rod, its velocity suddenly drops, as shown in Fig.3d. The selected two points in the plot were at the distance of x=0.625mm and x=5.0 mm along the interface (x=0 is the location of the first contact between the flyer and the inner rod, see Fig. 2b for the reference coordinate system). We can see that the $V_y$ reached a maximum at different times for various locations, with the progressive collision. The $V_y$ is defined as the velocity just prior to the sudden drop, and it was captured along the inside nodes of the tube along the welding direction. $\alpha$ was obtained by $\tan^{-1}(V_x/V_y)$ at the corresponding onset time of the CEMM simulation, as illustrated in Fig. 3c. The position where $V_y$ and $\alpha$ are extracted in the CEMM model corresponds exactly to the region where the specimen was taken for the experimental observations. In this work, in order to avoid the uneven distribution of Lorentz force at the slot side (marked in Fig. 2a) and its opposite side (180° to the slot), the position with a homogenous Lorentz force distribution (90° to the slot) was selected for analysis.

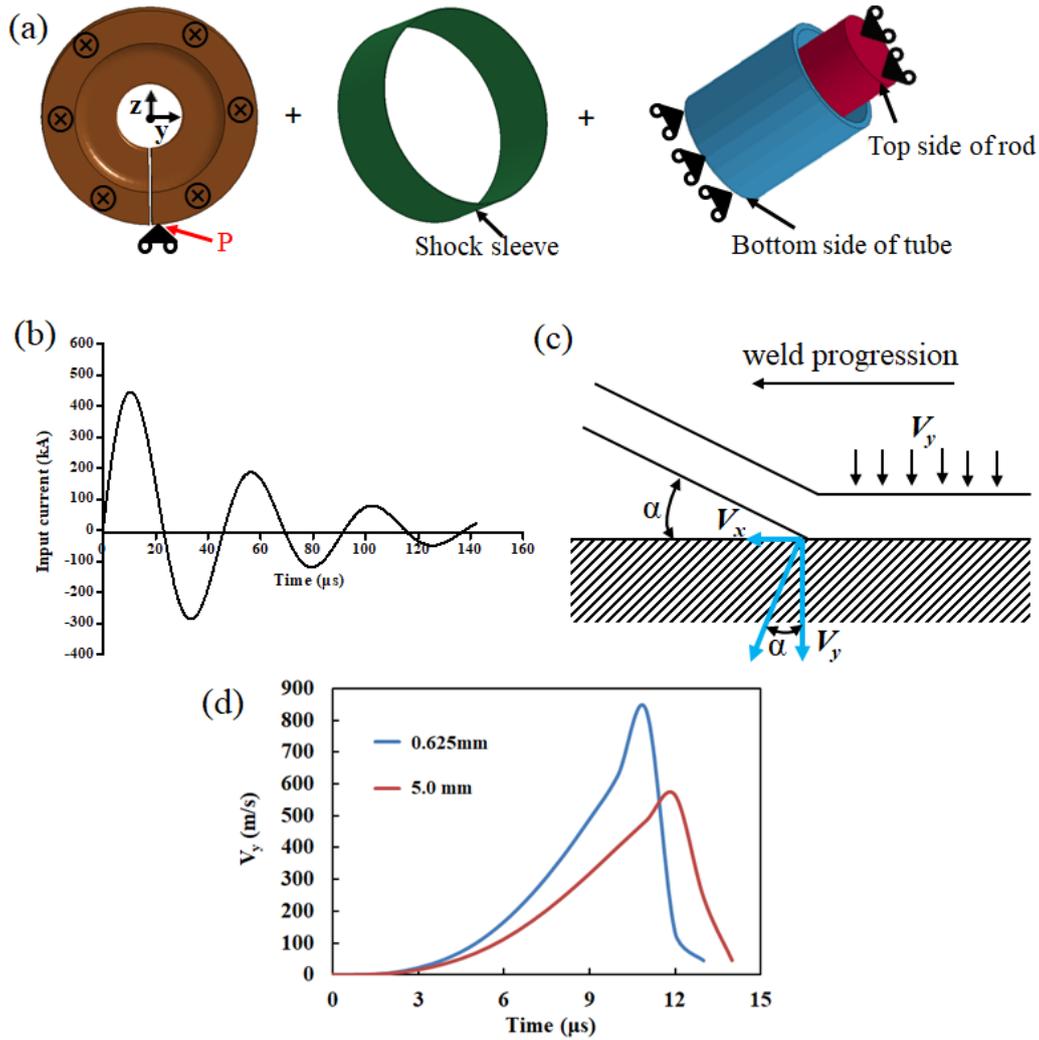

Fig. 3 (a) Boundary conditions and (b) input current used in the coupled electromagnetic-mechanical simulation; (c) Illustration of the angle measurement convention used in this study, where $V_x$ is the horizontal component of the impact velocity and $V_y$ is the vertical component of the impact velocity; (d) $V_y$ versus time curves obtained from CEMM model at two different locations.

## 3.2 Eulerian simulation

A thermomechanical model based on Eulerian formulation was constructed within ABAQUS® FE package to predict the modifications of the wave interface and to understand the fundamental mechanism and thermomechanical kinetics during the wave formation. The Eulerian model used a 2D equivalent geometry without activating the out-of-plane degree of freedom (DOF) and has the same size as the experimental welding parts, see Fig. 4a. The model consists of three sub-domains: Al flyer, Cu rod and the air gap. Appropriate boundary conditions are prescribed for both Al flyer and Cu rod, see Fig. 4b. The vertical component of impact velocity obtained from the CEMM simulation was applied to the flyer as an input to the Eulerian simulation (see the flowchart in Fig.1).

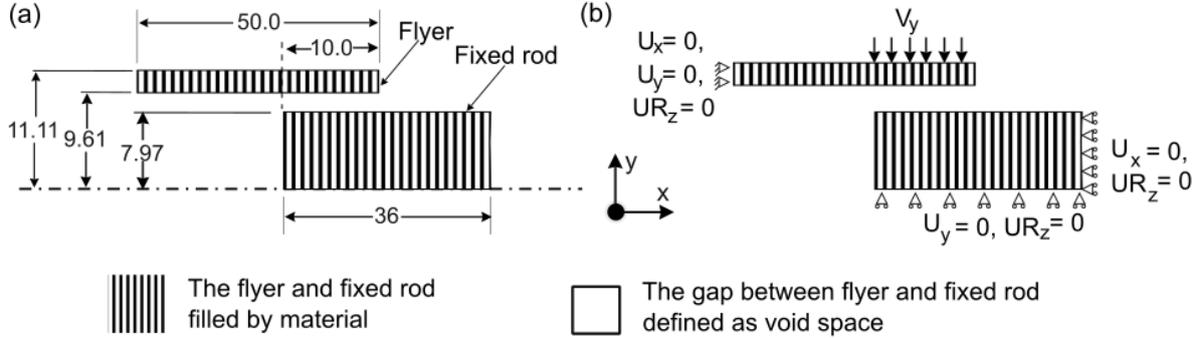

Fig. 4 (a) Schematic illustration showing geometrical details and (b) boundary conditions of the FE Eulerian model.

The Eulerian model was meshed with linear hexahedron, 8-node, EC3D8R elements, and it consists of 152613 elements with 305406 nodes. To improve the simulation accuracy and reduce the computational time, a finer mesh with the side length of 10 μm was used at the interfacial zone, while a coarser mesh was used at the zone away from the interface. The high strain-rate plastic behavior during the welding was captured using Johnson-Cook constitutive law (Gupta *et al.*, 2019) (given in Eq. (1)).

$$\sigma_y = [A + B(\varepsilon_p)^n]\left(1 + C \ln\left(\frac{\dot{\varepsilon}_p}{\dot{\varepsilon}_{p0}}\right)\right)(1 - T^{*m}) \qquad (1)$$

where, $\sigma_y$ is the yielding stress, $\varepsilon_p$ is the effective plastic strain, $\dot{\varepsilon}_p$ is the effective plastic strain rate, $\dot{\varepsilon}_{p0}$ is the reference strain rate, and $T^* = \frac{T-T_0}{T_m-T_0}$ is the dimensionless temperature. *A, B* and *C* are material parameters, *n* is the strain hardening exponent and *m* is the softening exponent.

Mie-Grüneisen's equation of state was used to compute the thermodynamic and pressure-density evolution during progressive collision. The temperature model used the energy equation considering the heat dissipation due to plastic work. Due to the extremely short period of impact process (less than 50 µs), the convective heat transfer may be neglected, and adiabatic boundary conditions were applied in the thermomechanical model.

The high strain rate material properties used in the Eulerian model is obtained from literature (Gupta *et al.*, 2006 and Gupta *et al.*, 2019). They are provided in the Supplementary Material (Table S2). In these works, the authors obtained the modulus of elasticity, Poisson's ratio and yield strength using uniaxial tensile test at a fixed strain rate. Moreover, they obtained the true strain and stress by using the Eqs. (2) and (3), respectively. Then, the stress-stain curve was used to compute the hardening coefficient *B* and strain hardening exponent *n* by using the least-squares method. The constant *C* and *m* are determined from Split-Hopkinson bar tests (Clausen et al., 2004).

$$\bar{\varepsilon}^{pl} = 2\ln(\frac{d_0}{d}) \qquad (2)$$

$$\frac{\bar{\sigma}_x}{\bar{\sigma}} = \left(1 + \frac{2R}{a}\right)\ln(1 + \frac{a}{2R}) \qquad (3)$$

## 4. Results and discussion

### 4.1 Impact kinematics along the interface

Large localized pressure waves created at the collision point during the welding process are travelling close to the velocity of sound in the material. Oppositely, the collision point progresses forward at a slower rate, i.e. subsonic rate (Akbari Mousavi and Al-Hassani, 2005). Therefore, the created pressures at the interface are sufficient to produce a strong shear instability which enables to generate a jetting phenomenon (Fig. 2c) at the interface (Nassiri *et al.,* 2016, Li *et al.*, 2020). According to Raoelison *et al.* (2012), this jetting phenomenon can remove the surface oxides and impurities from impacting surfaces and promote sound bonding. Moreover, one should note that the changes of the impact velocities and impact angles during the progressive collision can create various interface features for a given weld. Thus, the investigation of $V_y$ and $α$ along the welding interface is required to understand the physics behind the formation of various interface characteristics during MPW. Fig. 5 presents the velocity and angle along the interface obtained from the CEMM simulation. The distance (8.1 mm) finishing the black solid line in Fig. 5 indicates the regions where the Al flyer impacted the Cu rod during the simulation. The maximum velocity ($V_{max}$) and the simultaneously reached angle (*β)* given by the extended blue dashed line corresponding to the positions where the flyer does not impact the inner rod. From the CEMM simulation, we can clearly distinguish the non-impacted zone which does not have either a change in velocity direction or sharp drop in the magnitude. To differentiate them from the $V_y$, it is called $V_{max}$. The corresponding angle, *β*, for the non-impacted part of the weld is defined as $\tan^{-1}(V_x/V_{max})$ (for the definition of $V_x$ see Fig. 3c).

The curves indicate that $V_y$ and $α$ significantly fluctuate during the MPW process as the weld proceeds (Fig. 5). $V_y$ variation occurs mainly because the first impact occurs with the highest acceleration of the free edge of the flyer. Then the impact intensity decreases with decreasing acceleration along the weld progression as it reaches the fixed end. Moreover, one should note that the Lorentz force has a nearly homogeneous distribution throughout the welding section of the flyer. Thus, the spatial variation of the Lorentz force has a negligible influence on that of flyer velocity. It is worth noting that the highest velocity appears below the top side of the flyer tube slightly above the horizontal mid-plane of the fieldshaper (see Figs. 2a and b). The impact region can be further classified into Zone 1, Zone 2 and Zone 3, based on the characteristics of $V_y$ and $α,$ as marked in Fig. 5. At the onset of Zone 1, $V_y$ and $α$ fluctuate significantly which may not be favorable for welding. After that in Zone 2, $V_y$ starts to slightly decrease while $\alpha$ is rather stable, which is expected to produce a higher shearing instability at the interface. In Zone 3, $V_y$ gradually decreases in conjunction with the gradual increase of $\alpha$, which is expected to produce a less chaotic behavior for the interface instability. The variations in $V_y$ and $\alpha$ are expected to produce different interface features. Therefore, the corresponding classified interface zones of the experimental sample are further characterized to understand the consequence of the impact parameters on the interface behavior (Section 4.2).

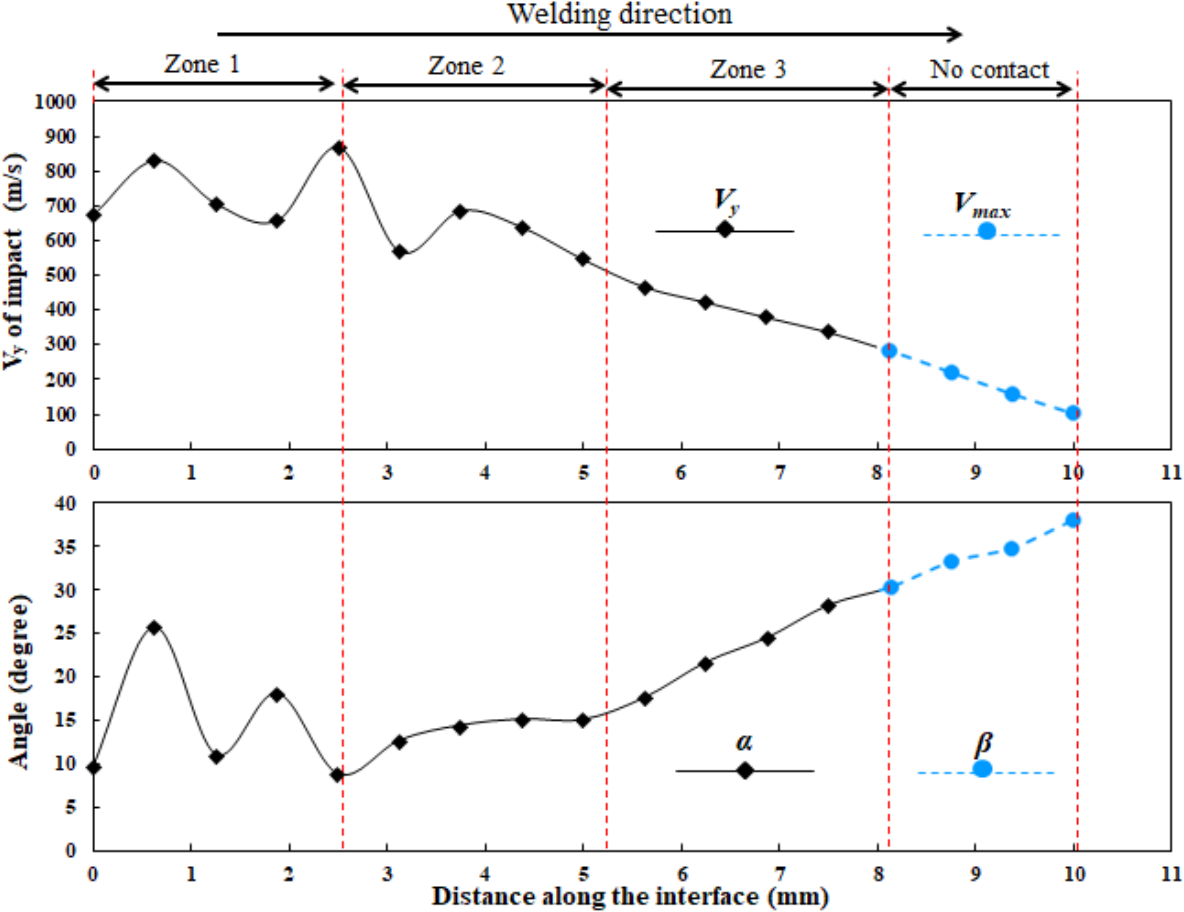

Fig. 5. Velocity and angle along the interface obtained from the CEMM simulation. The solid black lines portions of the curves (distance ≤ 8.1 mm) correspond to the $V_y$ of the flyer, which impacted the inner rod. Beyond the distance 8.1 mm, the curves are extended by blue dashed lines where it corresponding to the maximum velocity of the flyer reached during the process without contact to the inner rod.

### 4.2 Interface characterization

Fig. 6a shows the cross section of the MPW weld, revealing various interface morphologies along the welding direction (from right to left) corresponding to various impact conditions (Fig. 5). They include (1) unwelded zone at the onset of the weld (Fig. 6b); (2) formation of vortices + IM layers along the interface (Fig. 6c-e); and (3) various wave morphologies (with different amplitudes and wavelengths) (Figs. 6f-i). At the onset of welding, the impact conditions correspond to Zone 1 in Fig. 5, the higher $V_y$ with large fluctuation leads to produce a shock-like behavior, and thus the interface does not form a jet immediately. Cuq-Lelandais *et al.* (2016) showed that the reflected wave produces a tensile state at the collision point, enabling to peel off the interface and form the unwelded zone, even though the impact conditions are within the welding criterion. Psyk *et al.* (2017) confirmed this effect by numerical prediction for a sheet welding process. The flyer edge hits the target and is subsequently lifted. As welding proceeds (impact conditions corresponding to Zone 2 in Fig. 5), jetting phenomenon initiates due to severe interface shearing. This subsequently generates a strong confined heating which enables to melt the metals at the interface and form IM layers. Further detail of this Zone 2 and IMs are characterized in Section 4.3. Then, the interface instability decreases for the corresponding impact conditions in Zone 3 of Fig. 5. In Zone 3, the impact energy is insufficient to melt the base metals; and thus it forms either the wavy interface with a discontinuous thin (<20 μm) or even without an IM layer. One may note that at the end of our experimental welding sample, no contact zone is indeed observed at the interface (see Fig. 6a and Fig. S1(b) in the Supplementary Material). This is also in good agreement with our numerical prediction shown in Fig. 5 (indicated by the blue dashed lines in Fig. 5).

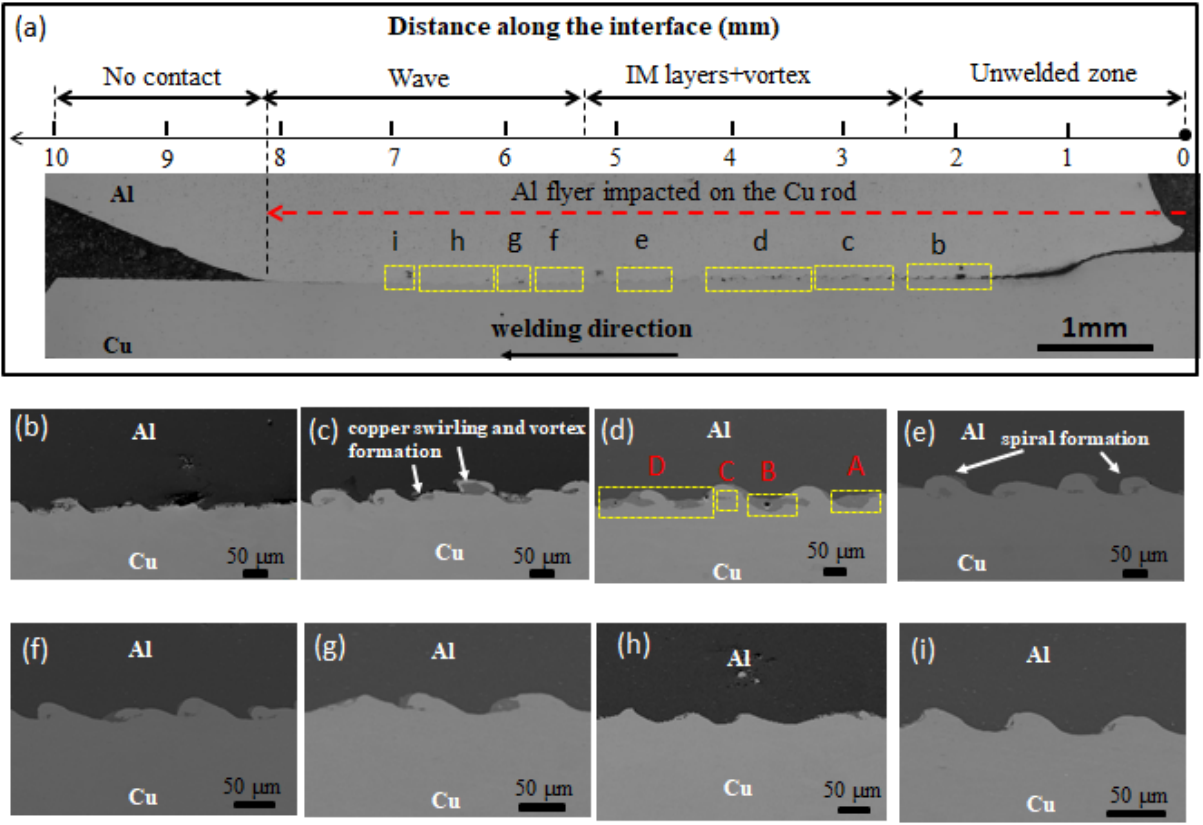

Fig. 6. (a) Cross-sectional micrograph of the MPW weld where various regions are marked by b-i, x axis represents the distance along the interface (mm); (b) unwelded zone at the onset of welding (corresponding to zone 1 in Fig. 5); (c)-(e) interface with IM layers + vortices (corresponding to Zone 2 in Fig. 5); (f) - (i) wavy interface (corresponding to Zone 3 in Fig. 5). (b)-(i) correspond to the large magnification images of the areas marked by 'b-i' in (a), respectively.

**4.3 Microstructural characterization**

Fig. 7 shows the detailed SEM images of the interfacial area shown in Fig. 6d, revealing complex interface morphologies within the IM layers. The thicknesses of the IM layers are varying from 1.2 μm to 37.6 μm along the interface. Fig. 7a shows a clear mixing pattern of the spiraling trajectories in the IM layers. As the collision point advances, a swirling of the materials was induced at the interface due to the shear instability. Therefore, the interface materials twist and roll up like in a fluidic interface to form vortex. This phenomenon is usually ascribed to the well-described fluid mechanics Kelvin-Helmholtz instability caused by a shearing instability across the interface of two fluids. According to Li *et al* (2020b), when the interface instability reaches advanced stages, the materials within the vortex experiences large velocity gradients and causes the swirling flow.

Pores with micron/submicron size also appear within the IM layers. When comparing their size, we can classify them into two types, i.e., pore diameters smaller than 1.0 μm (submicron pores) (marked by red arrows in Fig. 7a-d) and microscale pores with diameters between 2.8 μm and 15 μm (marked by green arrows in Fig. 7). Moreover, the submicron pores present a random distribution while the microscale cavities always appear in the middle of the IM layers. This indicates that their formations are governed by different physical mechanisms. During the MPW, the interface materials experienced a sudden increase in temperature followed by a rapid cooling under isochoric conditions. This phenomenon provides a favorable condition for the formation and growth of submicron sized pores. These pores are inevitably formed as they ensure the conservation of the total volume in the materials, as reported in our previous works (Sapanathan *et al*. (2017) and Raoelison *et al.* (2019)). During the porous structure formation, the pressure gradient in the concerned zone promotes the depressurization essential for cavitation. As suggested by Hamada *et al.* (2007), the depressurization within the molten liquid zone changes the surface tension, and results in the fluid rupture providing the required condition for the nucleation of the fine pores. Concurrently, contraction/shrinkage occurs during cooling, which further facilitates the growth and coalescence of the pores within the confined volume.

The microscale cavities are spherical as evidenced in Fig. 7. It demonstrates that the microscale cavities result from physical kinematics, instead of purely based on the high impact pressure that would expectedly shrink or distort such cavities. As welding progresses, the interface can trap an empty space when it experiences ultra-high impact. Concurrently, local melting results in the formation of a liquid phase. A very fast circular motion occurs in the liquid phase at the vicinity of the trapped empty space due to the different shearing between the back and front sides of the pore. This fast circular motion enables the irregular shape of the trapped empty space to transform into a circular one. Thus, the mixing of molten zone combined with an intense swirling result in the formation of microscale cavities within the vortex zone. According to Li *et al*. (2020b), the combination of the confined local melting, thermomechanical softening and ultra-high cooling provide a favorable condition to freeze the pores within the intermediate layer.

Some microcracks are also observed inside the thick zones of solidified melted IM layers (indicated by white arrows in Figs. 7a, b and d). They are bounded within the IM layers and never propagate inside the Al or Cu. As discussed by Sapanathan *et al.* (2019), these microcracks are resulting from thermal residual stress induced by the temperature gradient between the solidified melted and the parent metals during the solidification shrinkage of the welded structure. In Fig. 7c and d, some fragments (marked by the yellow arrows) are also found in IM zones. In order to understand the chemical compositions of the IM zones, EDS maps of Al and Cu are further analyzed within a selected IM zone (marked by red rectangle in Fig. 7d). Fig. 7e and f show the EDS maps for Al and Cu respectively, and Fig. 7g shows the overlaid Al and Cu maps. Fig. 7h shows the quantitative analysis of weight fraction of Al and Cu corresponding to the squares A, B and C (marked in Fig. 7g). It confirms the fragments are Cu and intermetallic compound is $Al_2Cu$ in the IM layers according to the atomic percentage.

According to Wang *et al.* (2020), the combined formation of the liquid/solid interface within the IM zone, the high pressure and high temperature could promote the fast reaction and diffusion kinetics of Cu in Al liquid, and results in the formation of $Al_2Cu$.

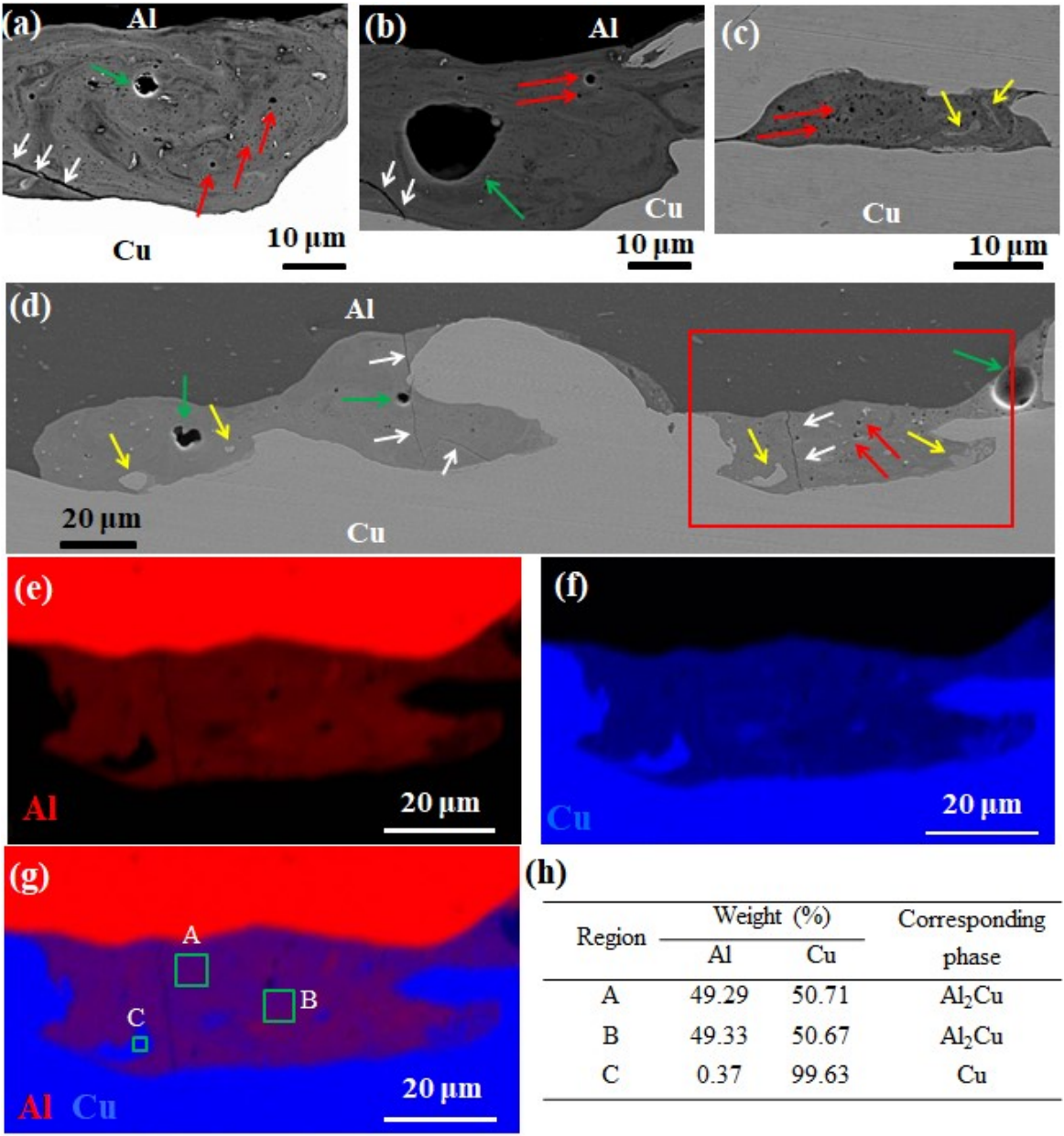

Fig. 7. Microstructures of the IM layers showing (a) swirling and porous structures; (b) microscale cavity; (c) adjacent intermixing zone with micro porous structures; (d) microcracks bounded in the IM layers; (a), (b), (c) and (d) are taken from the areas A, B, C and D marked in Fig. 6d, respectively. The microscale pores, microcracks and Cu fragments are highlighted by green, white and yellow arrows, respectively. EDS analysis: (e) Elemental map of Al, (f) Elemental map of Cu, (g) Elemental map of Al and Cu, (h) EDS results for the square regions corresponding to A, B and C in Fig. 7g.

**4.4 Wave formation**

**4.4.1 Numerical predictions of the wave formation**

The physically realistic descriptions of the wave formation involve complex kinematics which are not possible to capture via *insitu* experimental methods due to the highly dynamic and transient impact conditions. Therefore, the obtained impact conditions ($V_y$) from CMME simulation (Fig. 5) are used as input for the Eulerian model (see section 3.2) to further investigate the wave formation. The used portion of the input velocity corresponding to the wave formation (Zone 3 in Fig. 5 and Figs. 6f-i) can be described by Eq. (4).

$$|V_y| = -300 - 70000x; \quad \forall x \in [0, 0.0026] \quad (4)$$

where, $V_y$ is the vertical component of the impact velocity (Unit: m/s), and $x$ is the distance from the end to the onset of the wave formation (Unit: m).

The images in Figs. 8a-d show various wave morphologies obtained from the Eulerian simulation limited to Zone 3 of Fig 5 and 6. The simulation results corroborate the experimental observations as shown in Figs. 8i-l, in terms of the shape, size and sequential development of wave morphologies. This indicates that the thermomechanical model is a reliable tool to further analyze the formation mechanisms of the wave which will be discussed in the following sections. As we can see, the interface kinematics (wave morphologies and vortex formation) are governed by the local impact conditions (impact velocity and impact angle). To investigate the governing mechanisms of the complex interfacial morphologies, it is first required to validate the experimentally observed wave morphologies using the Eulerian computation. After the validation of the Eulerian model based on morphology (by comparing the predictions to the experimental observations), the data is further post-processed to inversely investigate the underlying potential mechanisms that governs the particular interface formation. Moreover, it should be noted that for the Eulerian simulation, one model cannot predict all the wave morphologies. In this study, different output frequencies were used to get the different local condition (collision point velocity) to produce different types of waves (Fig. 8).

The simulation also enables to investigate the temperature distribution along the wave interface (Figs. 8e-h), impossible to measure by *insitu* methods during this high-speed transient welding process. The melting point of Al (~660 °C) is set to the limit temperature for the contours. The temperature distribution maps show a narrow band with significantly excessive heating along the wavy pattern at all locations. The localized gray pockets shown in Figs. 8e and f indicate that the temperature reached is higher than 660 °C, and even up to 1279 °C, which exceeds the melting point of Cu (~ 1085 °C). Those zones with the presence of high temperature, are in good agreement with the IM phases in the experimental observation (in terms of shape and site occurrence), and it indicates that a possible local melting facilitate the formation of IM phases (Figs. 8i and j). In contrast, the predicted distributions of temperature corresponding to the Type 3 and 4 waves shown in Figs. 8g and h, are below the melting point of both Al and Cu. It also concurs with the experimentally observed waves with the absence of IM phases at the interface. Based on these observations, we could conclude that the wavy interfaces experience two bonding mechanisms, (i) mixture of local melting and solid-state bonding (for Type 1 and 2) and (ii) solid-state bonding alone (for Type 3 and 4). The observed bonding mechanisms were also reported in the recent work in Psyk *et al.* (2019) for dissimilar MPW of sheet metals using experimental observations.

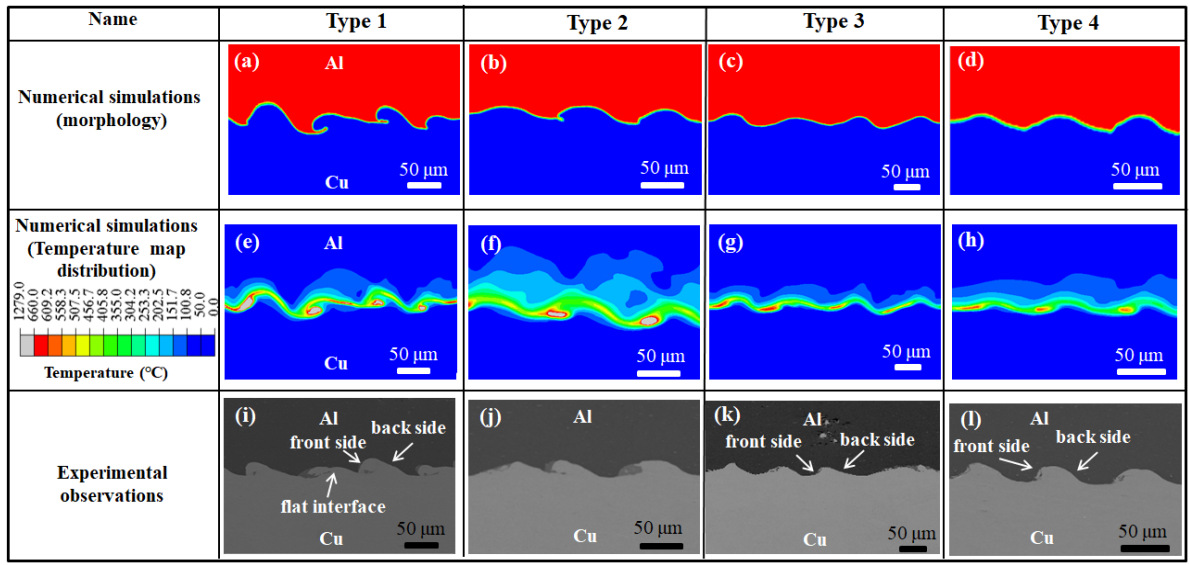

Fig. 8. Various wave morphologies obtained from the numerical simulations (a-d), corresponding temperature distributions (e-h) and the experimental observations of the waves (i-l).

### 4.4.2 Average equivalent plastic strain at the interface

The contour plot of average equivalent plastic strain obtained from the Eulerian simulation is shown in Figs. 9a-d. It is computed as a volume fraction weighted average of all materials in the finite element. The result reveals a confined plastic deformation along the interface in all types of waves as, which well explains the temperature rise of the interface (Figs. 8e-h). The plastic strain distribution clearly indicates the higher plastic values at the vicinity of the interface, while the plastic strain rapidly decreases to zero within the region near to the base metals. Moreover, the regions corresponding to higher strain values are also in good agreement with the IM phase formation (Fig. 8i and j). This reveals that the average equivalent plastic strain could also be a criterion for predicting the formation of wavy interface with or without the IM phase. For specific material combinations, the plastic strain can be used to inversely predict the impact conditions which further enable to identify the process parameters. To establish a welding window based on the plastic strain criterion, it requires multiple simulations and inverse analysis of the local strain to identify the corresponding process parameter.

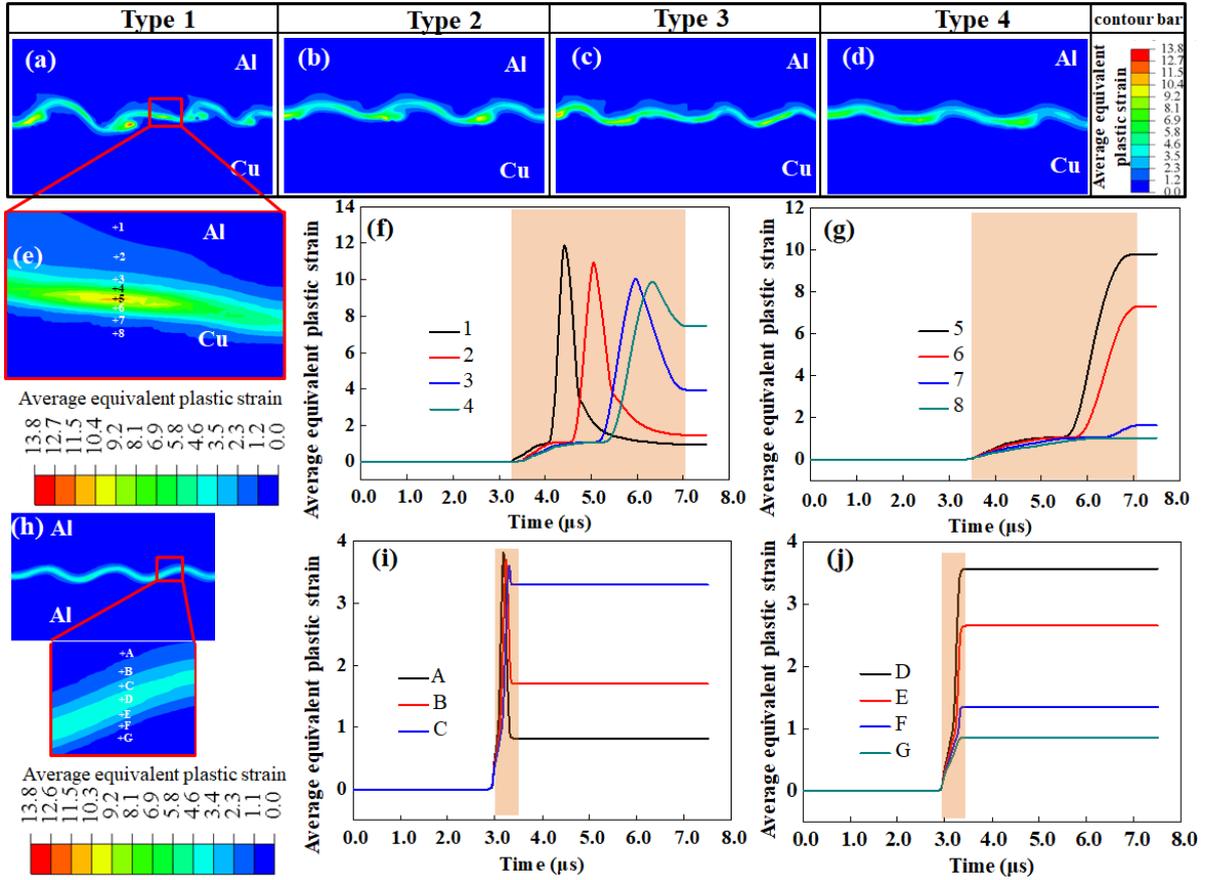

Fig. 9. (a), (b), (c) and (d) represent the simulation results of the average equivalent plastic strain distribution at the wavy interface for the wave Type 1, 2, 3 and 4, respectively; (e) A magnified average equivalent plastic strain map of the area marked by red rectangle in wave Type 1 in (a); changes of average equivalent plastic strain over time at points on the (f) flyer side and (g) rod side near the interface marked in (e); (h) average equivalent plastic strain map obtained from Al/Al interface with the same impact conditions of Type 1 wave; changes of average equivalent plastic strain over time at points for (i) the Al flyer and (j) the Al rod at an Al/Al interface marked in (h). The regions highlighted by orange color in Figs. 9f, g, i and j represent the plastic strain evolution during wave formation.

A magnified plastic strain image of wave Type 1 (Fig. 9e) illustrates the position of points 1-4 corresponding to the Al flyer and points 5-8 corresponding to the Cu rod near the interface. These points were chosen to plot the temporal evolution of the average equivalent plastic strain, as shown in Figs. 9f and g for Al and Cu sides, respectively. When t=0, it denotes the beginning of the flyer impact onto the inner rod. As welding propagates to the selected regions the plastic strain starts to substantially increase. The point of initial plastic strain rise occurs at t=3.4 μs (Figs. 9f and g) and t=2.8 μs (Figs. 9i and j) for Al/Cu and Al/Al case, respectively. After that time, the plastic strain evolves until the impact passes the selected region. Therefore, a region marked by shaded orange color in Figs. 9f, g, i and j has been added to highlight the times which corresponding to the plastic strain evolution during the wave formation. The results show that all the points on the flyer side (points 1-4) experience a rise in plastic strain, and then the strain levels drop to specific values which remain at the interface (Fig. 9f). However, all the plastic strain curves on the Cu rod (points 5-8) show that the strain values increase to the maximum and then directly remain inside the material (Fig. 9g). At the end of the plastic strain curves, the strain values do not drop down to zero, it means some residual plastic strain remains inside the material after the wave formation. To further examine this difference, we also investigate the average equivalent plastic strain for a reference Al/Al interface as shown in Fig. 9h. A similar trend for the strain curves of the Al/Cu interface is also observed at the Al/Al interface (Fig. 9i and Fig. 9j corresponding to the Al flyer and Al rod, respectively). This result confirms that the "peak" values in the strain curves on the flyer side (Fig. 9f) are not due to the different flow behaviors between the Al and Cu. Instead, it is attributed to the formation of waves resulting from the repeated deformation instead of immediately being formed after the first onset of impact.

### 4.4.3 Wave morphology and jetting kinematics

To further study the kinematics of the wave formation, three typical wave morphologies are chosen, i.e. cases where (i) the wavelength on the front side is shorter than on the back side (Type 1), (ii) the wavelength on the front side is longer than on the back side (Type 3) and (iii) the wavelength on the front and back side are approximately equal (Type 4) (front side and back side are marked in Figs. 8i, k and l). Since it is difficult to distinguish the original surface of both Al and Cu after the welding process, the middle of the highest and lowest vertical displacement of each wave type is considered as the original welding surface. The vertical displacements obtained for waves Type 1, 3 and 4 are depicted in Figs. 10a, b and c respectively. For Type 3 and 4 waves, their vertical displacements (Fig. 10 b and c) increase with increasing $\alpha$. Watanabe and Kumai (2009) studied dissimilar Al-Cu welds and showed that, along the welding direction, the welding interface presents highly non-uniform wavelength and amplitude compared with the gradual variation in the Al/Al and Cu/Cu welds. It indicates that more complex interfacial kinematics are present during the wave propagation at the Al/Cu interface.

Raoelison *et al.* (2016) suggested that, during the MPW process, the upward and downward jetting contribute to the wave morphologies. More details of the wave development are provided in supplementary material (see Fig. S2, Fig. S3 and Fig. S4). Figs. 10d, e and f show the corresponding jetting angles along the waves Type 1, 3 and 4, respectively. The inset figure in Fig. 10f defines the jetting angle calculation from the simulation. The results reveal an excellent agreement between the wave morphologies obtained by image analysis using micrographs (Figs. 10a-c) and the jetting angle morphologies obtained from the Eulerian simulation (Figs. 10d-f). The flat interface shown in wave Type 1 (between the distance along the interface of 5.65 mm and 5.70 mm in Fig. 10 a and d, marked in Fig. 8i), is due to the approximately zero jetting angle during the wave development. It should be noted that the wave Type 1 is more irregular compared to the waves Type 3 and 4 in terms of the wavelength and amplitude. This could be explained by the values of the jetting angles (Fig. 10d) which is sometimes higher than the impact angle (Fig. 5) during the impact process.

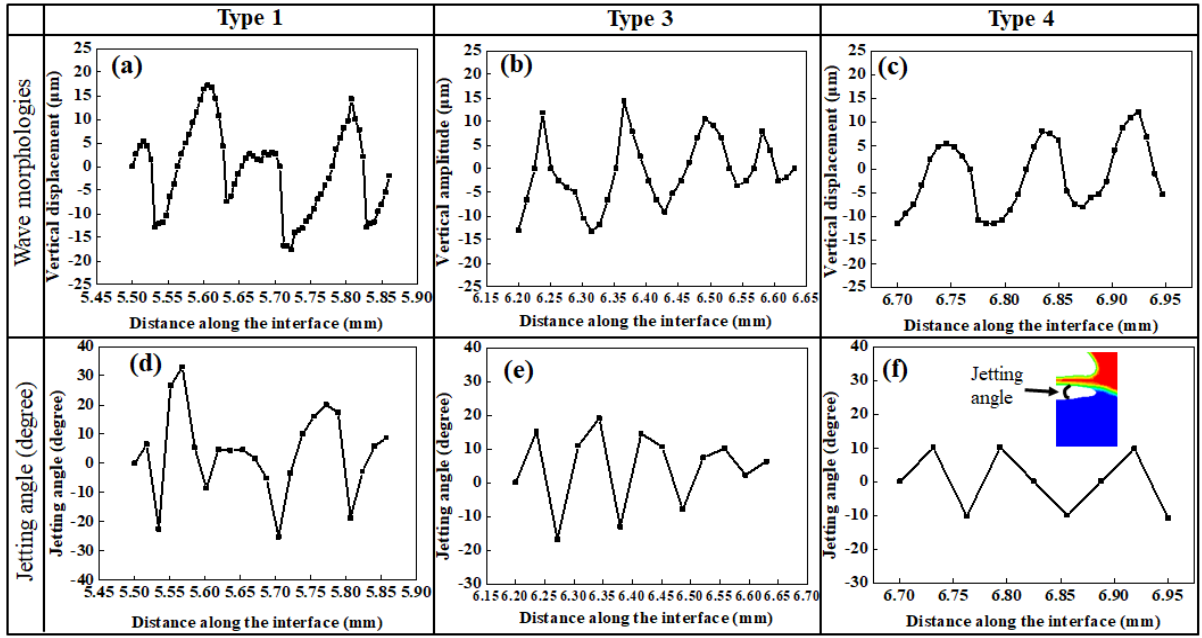

Fig. 10. (a), (b) and (c) present wave morphologies of Type 1, 3 and 4 obtained by image analysis using micrographs; (d), (e) and (f) are the predicted jetting angle from Eulerian simulation corresponding to wave Type 1, 3 and 4, respectively. The inset in (f) explains the jetting angle calculation from the simulation.

In Fig. 10, we find a very good correlation between the jetting angle and corresponding displacement at a given instant; i.e. larger jetting angle produces larger vertical displacement. This effect is further illustrated in Fig. 11, which demonstrates that the absolute values of jetting angle and the vertical displacement exhibit an approximately linear relationship. Since a higher shear instability and larger pressure can produce higher jetting angle, we can thus expect that a large shear instability could produce waves with larger amplitude which also have a high probability to form IM phases.

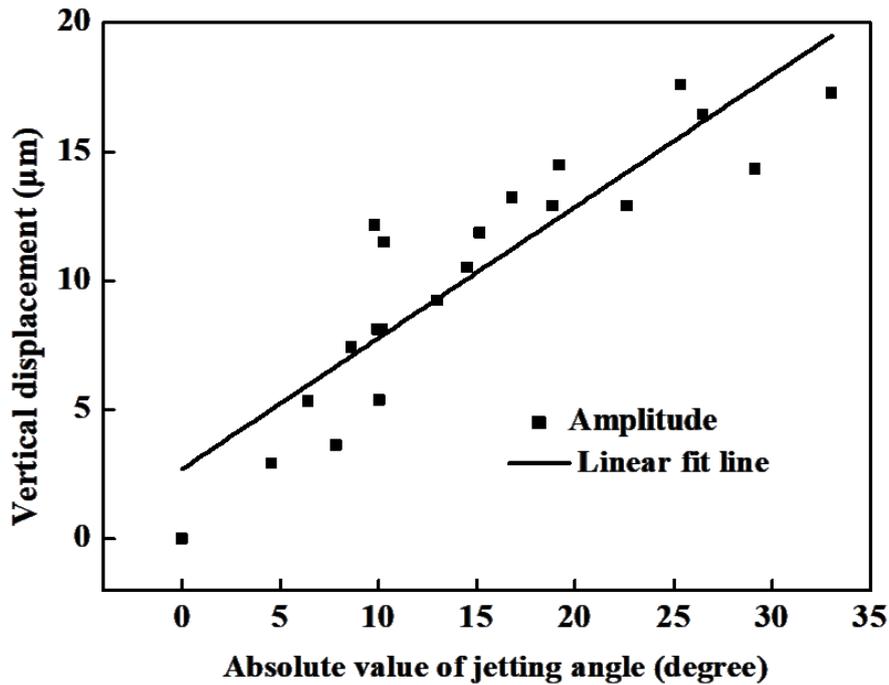

Fig. 11. Correlation between the jetting angle and the amplitude obtained from the waves.

**4.4.4 Collision pressure and collision point velocity**

The pressure at the collision zone is an important parameter during the wave formation. It must be high enough to exceed the dynamic yield strength of the base metal to produce a jet. Since the collision zone reveals highly dynamic variation (see an example of predicted pressure distribution in Fig. S5 in the supplementary material), the average pressure at the collision zone is plotted in Fig. 12a. It can be observed that the pressure generally decreases from the onset to the end for each wave while having some occasional spikes. Compared with the pressure, the collision point velocity experiences more fluctuations, as shown in Fig. 12b. However, the collision point velocity exhibits the same trend as the collision pressure.

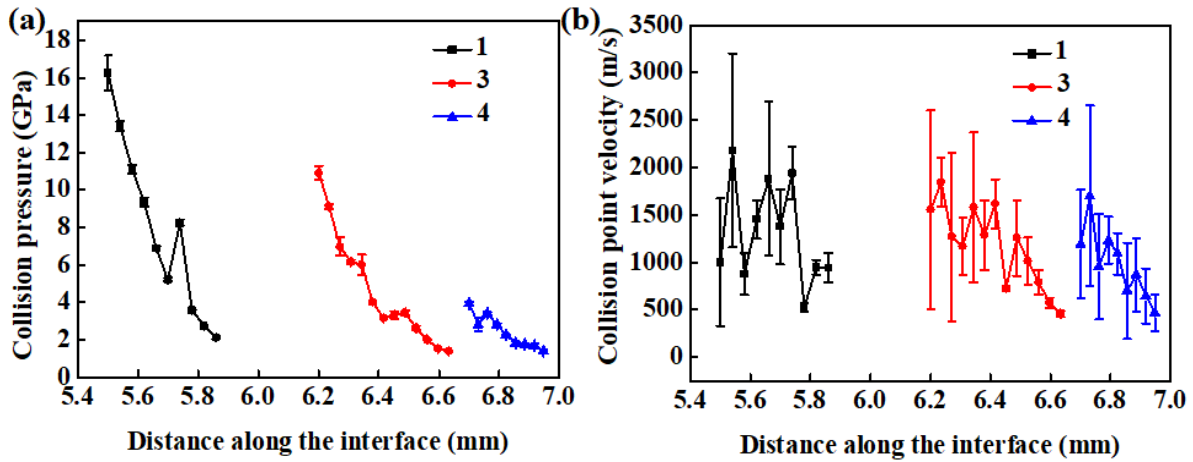

Fig.12. Collision pressure and collision point velocity along the wave interface obtained from Eulerian simulation for type 1, 3 and 4 waves (Fig. 10). Error bars represent the standard deviation, where the average and standard deviation of the pressure and velocity are calculated using 9 points from the simulation (the point with max pressure and the surrounding 8 nodes in a quadrilateral mesh).

Since the collision point velocity is almost parallel to the interface, it could predominantly affect the wavelength according to Wang *et al.* (2020). In this study, the half wavelength of each wave was measured in order to get more reliable data since the waves are not symmetrical. The half wavelength increases with increasing collision point velocity, see Fig. 13a. However, the impact velocity also affects the wave propagation, which may inversely affect the wavelength. The relationship between the ratio of "collision point velocity/impact velocity" and the half wavelength is depicted in Fig. 13b. Although Fig. 13b reveals a similar trend as in Fig. 13a, the curve fitting confidence is higher in the case of Fig. 13b meaning the ratio is more suitable as an influencing factor for evaluating wavelength during the MPW process.

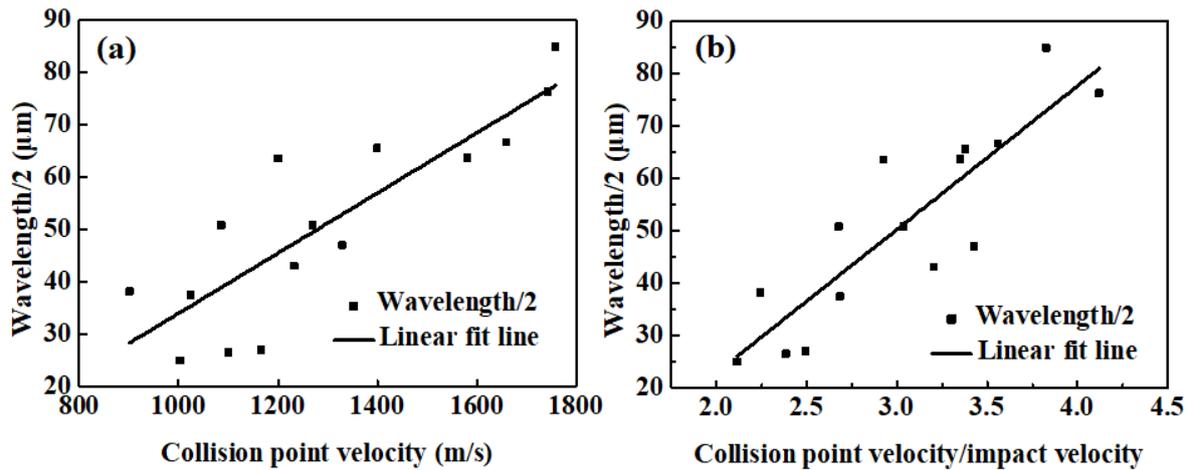

Fig. 13. Half wavelength evolution with (a) the collision point velocity and (b) with the ratio of "collision point velocity/impact velocity".

5 Conclusions

A complex Al/Cu joint with various interfacial morphologies obtained from magnetic pulse welding (MPW) was systematically investigated using experimental characterizations and numerical simulations. The important findings of this study can be summarized as follows:

(1) Coupled electromagnetic mechanical (CEMM) simulations are used to identify the local impact conditions during the MPW. Various welding zones were classified based on the prediction of impact velocity and impact angle. Those classified welding zones based on the impact conditions well correlate to the experimentally observed unwelded zone, vortex + IM zone, wavy interface and no contact zone along the welding direction.

(2) Continuous IM layers are characterized by highly heterogeneous porous zone with random size and distribution due to a cavitation phenomenon. Microscale cavities in the center of the continuous IM layers are formed due to the combination of trapped empty space and drastic swirling under sufficient heat and ultra-fast cooling conditions. Continuous IM layers are formed by mechanical mixing combined with melting.

(3) A coupled electromagnetic-mechanical simulation demonstrates that the various

morphologies described in point (1) and (2) of this conclusion appear due to the highly dynamic impact velocity and impact angle along the interface. The impact velocity and impact angle exhibit significant fluctuations at the onset of welding. Then, the impact velocity gradually decreases with the increase of the impact angle during the propagation of welding.

(4) An Eulerian simulation is used to successfully reproduce the various experimentally observed waves in terms of morphology and site occurrence. The temperature distribution and average equivalent plastic strain along the interface well correlate with the experimentally observed IM phases due to local melting and severe plastic deformations in the corresponding zones. The actual kinematics during the wave generation demonstrates that the wave morphology is formed with repeated deformations of the interface rather than with a single-step deformation.

(5) A larger jetting angle enables to increase the vertical displacement of the waves and produces large wave amplitudes. Although both collision point velocity as well as the ratio between collision point velocity and impact velocity exhibit strong influence on the wavelength, the latest ratio is the most suitable for evaluating the wavelength due to the increased confidence in the sampling distributions.

**Acknowledgements**

This work is funded by the China Scholarship Council (No. 201701810138). Jishuai Li acknowledges the financial support from the Doctoral School of UTC during his research stay at UCLouvain in 2019. T. Sapanathan acknowledges F.R.S–FNRS (Belgium) during his postdoc at UCLouvain. Yuliang Hou acknowledges the financial support from National Natural Science Foundation of China (NSFC, No, 52005451).

# References

Akbari Mousavi, A.A., Al-Hassani, S.T.S., 2005. Numerical and experimental studies of the mechanism of the wavy interface formations in explosive/impact welding. J. Mech. Phys. Solids 53, 2501–2528. https://doi.org/10.1016/J.JMPS.2005.06.001

Acarer, M., 2012. Electrical, corrosion, and mechanical properties of aluminum-copper joints produced by explosive welding. J. Mater. Eng. Perform. 21, 2375–2379. https://doi.org/10.1007/s11665-012-0203-6

Avettand-Fènoël, M.-N., Simar, A., 2016. A review about Friction Stir Welding of metal matrix composites. Mater. Charact. 120, 1–17. https://doi.org/https://doi.org/10.1016/j.matchar.2016.07.010

Bahmani, M.A., Niayesh, K., Karimi, A., 2009. 3D Simulation of magnetic field distribution in electromagnetic forming systems with field-shaper. J. Mater. Process. Technol. 209, 2295–2301. https://doi.org/10.1016/j.jmatprotec.2008.05.024

Bataev, I.A., Tanaka, S., Zhou, Q., Lazurenko, D.V., Junior, A.M.J., Bataev, A.A., Hokamoto, K., Mori, A., Chen, P., 2019. Towards better understanding of explosive welding by combination of numerical simulation and experimental study. Mater. Des. 169, 107649. https://doi.org/10.1016/J.MATDES.2019.107649

Clausen, A.H., Børvik, T., Hopperstad, O.S., Benallal, A., 2004. Flow and fracture characteristics of aluminium alloy AA5083-H116 as function of strain rate, temperature and triaxiality. Mater. Sci. Eng. A 364, 260–272. https://doi.org/10.1016/j.msea.2003.08.027

Cui, J., Sun, G., Li, G., Xu, Z., Chu, P.K., 2014. Specific wave interface and its formation during magnetic pulse welding. Appl. Phys. Lett. 105, 1–5. https://doi.org/10.1063/1.4903044

Cuq-Lelandais. J. P, Avrillaud, G., Ferreira, S., Mazars, G., Nottebaert, A., Teilla, G., Shribman, V. 2016. 3D Impacts Modeling of the Magnetic Pulse Welding Process and Comparison to Experimental Data. 7th International Conference on High Speed Forming. 13–22.

Cui, J., Li, Y., Liu, Q., Zhang, X., Xu, Z., Li, G., 2019. Joining of tubular carbon fiber-reinforced plastic/aluminum by magnetic pulse welding. J. Mater. Process. Technol. 264, 273–282. https://doi.org/https://doi.org/10.1016/j.jmatprotec.2018.09.018

Faes, K., Kwee, I., De Waele, W., 2019. Electromagnetic pulse welding of tubular products: Influence of process parameters and workpiece geometry on the joint characteristics and investigation of suitable support systems for the target tube. Metals (Basel). 9. https://doi.org/10.3390/met9050514

Gupta, N.K., Iqbal, M.A., Sekhon, G.S., 2006. Experimental and numerical studies on the behavior of thin aluminum plates subjected to impact by blunt- and hemispherical-nosed projectiles. Int. J. Impact Eng. 32, 1921–1944. https://doi.org/10.1016/J.IJIMPENG.2005.06.007

Groche, P., Becker, M., Pabst, C., 2017. Process window acquisition for impact welding processes. Mater. Des. 118, 286–293. https://doi.org/10.1016/j.matdes.2017.01.013

Gupta, V., Lee, T., Vivek, A., Choi, K.S., Mao, Y., Sun, X., Daehn, G., 2019. A robust process-structure model for predicting the joint interface structure in impact welding. J. Mater. Process. Technol. 264, 107–118. https://doi.org/10.1016/J.JMATPROTEC.2018.08.047

Hamada, Y., Koga, K., Tanaka, H., 2007. Phase equilibria and interfacial tension of fluids confined in narrow pores. J. Chem. Phys. 127. https://doi.org/10.1063/1.2759926

Hahn, M., Weddeling, C., Lueg-Althoff, J., Tekkaya, A.E., 2016. Analytical approach for magnetic pulse welding of sheet connections. J. Mater. Process. Technol. 230, 131–142. https://doi.org/https://doi.org/10.1016/j.jmatprotec.2015.11.021

Kakizaki, S., Watanabe, M., Kumai, S., 2011. Simulation and experimental analysis of metal jet emission and weld interface morphology in impact welding. J. Japan Inst. Light Met. 61, 328–333. https://doi.org/10.2464/jilm.61.328

Kapil, A., Sharma, A., 2015. Magnetic pulse welding: an efficient and environmentally friendly multi-material joining technique. J. Clean. Prod. 100, 35–58. https://doi.org/10.1016/J.JCLEPRO.2015.03.042

Kwee, I., Psyk, V., Faes, K., 2016. Effect of the Welding Parameters on the Structural and Mechanical Properties of Aluminium and Copper Sheet Joints by Electromagnetic Pulse Welding. World J. Eng. Technol. 04, 538–561.

Lysenko, D.N., Ermolaev, V.V., Dudin, A.A., 1970. Method of pressure welding. Patentschrift US 3,520,049.

Lee, T., Zhang, S., Vivek, A., Daehn, G., Kinsey, B., 2019. Wave formation in impact welding: Study of the Cu–Ti system. CIRP Ann. 68, 261–264. https://doi.org/10.1016/j.cirp.2019.04.058

Li Z, Beslin E, den Bakker AJ, Scamans G, Danaie M, Williams CA, Assadi H, 2020. Bonding and microstructure evolution in electromagnetic pulse welding of hardenable Al alloys, J. Mater. Process. Technol. doi:https://doi.org/10.1016/j.jmatprotec.2020.116965

Li, J.S., Sapanathan, T., Raoelison, R.N., Zhang, Z., Chen, X.G., Marceau, D., Simar, A., Rachik, M., 2019. Inverse prediction of local interface temperature during electromagnetic pulse welding via precipitate kinetics. Mater. Lett. 249, 177–179. https://doi.org/10.1016/J.MATLET.2019.04.094

Li, J S, Raoelison, R.N., Sapanathan, T., Hou, Y.L., Rachik, M., 2020a. Interface evolution during magnetic pulse welding under extremely high strain rate collision: mechanisms, thermomechanical kinetics and consequences. Acta Mater. 195, 404-415. https://doi.org/https://doi.org/10.1016/j.actamat.2020.05.028

Li, J. S., Raoelison, R.N., Sapanathan, T., Zhang, Z., Chen, X.G., Marceau, D., Hou, Y.L., Rachik, M., 2020b. An anomalous wave formation at the Al/Cu interface during magnetic pulse welding. Appl. Phys. Lett. 116, 161601. https://doi.org/10.1063/5.0005299

Lueg-Althoff, J., Bellmann, J., Gies, S., Schulze, S., Tekkaya, A.E., Beyer, E., 2018. Influence of the flyer kinetics on magnetic pulse welding of tubes. J. Mater. Process. Technol. 262, 189–203. https://doi.org/10.1016/J.JMATPROTEC.2018.06.005

Nassiri, A., Kinsey, B., Chini, G., 2016. Shear instability of plastically-deforming metals in high-velocity impact welding. J. Mech. Phys. Solids 95, 351–373. https://doi.org/10.1016/J.JMPS.2016.06.002

Rice, J.R., Tracey, D.M., 1969. On the ductile enlargement of voids in triaxial stress fields. J. Mech. Phys. Solids 17, 201–217. https://doi.org/10.1016/0022-5096(69)90033-7

Psyk, V., Risch, D., Kinsey, B.L., Tekkaya, A.E., Kleiner, M., 2011. Electromagnetic forming—A review. J. Mater. Process. Technol. 211, 787–829. https://doi.org/10.1016/J.JMATPROTEC.2010.12.012

Psyk, V., Scheffler, C., Linnemann, M., Landgrebe, D., 2017. Manufacturing of hybrid aluminum copper joints by electromagnetic pulse welding - Identification of quantitative process windows. AIP Conf. Proc. 1896. https://doi.org/10.1063/1.5008128

Psyk, V., Hofer, C., Faes, K., Scheffler, C., Scherleitner, E., 2019. Testing of magnetic pulse welded joints - Destructive and non-destructive methods. AIP Conf. Proc. 2113. https://doi.org/10.1063/1.5112574

Raoelison, R.N., Buiron, N., Rachik, M., Haye, D., Franz, G., 2012. Efficient welding conditions in magnetic pulse welding process. J. Manuf. Process. 14, 372–377. https://doi.org/10.1016/J.JMAPRO.2012.04.001

Raoelison, R.N., Buiron, N., Rachik, M., Haye, D., Franz, G., Habak, M., 2013. Study of the elaboration of a practical weldability window in magnetic pulse welding. J. Mater. Process. Technol. 213, 1348–1354. https://doi.org/10.1016/j.jmatprotec.2013.03.004

Raoelison, R.N., Sapanathan, T., Buiron, N., Rachik, M., 2015. Magnetic pulse welding of Al/Al and Al/Cu metal pairs: Consequences of the dissimilar combination on the interfacial behavior during the welding process. J. Manuf. Process. 20, 112–127. https://doi.org/10.1016/j.jmapro.2015.09.003

Raoelison, R.N., Sapanathan, T., Padayodi, E., Buiron, N., Rachik, M., 2016. Interfacial kinematics and governing mechanisms under the influence of high strain rate impact conditions: Numerical computations of experimental observations. J. Mech. Phys. Solids 96, 147–161. https://doi.org/10.1016/j.jmps.2016.07.014

Raoelison, R.N., Li, J., Sapanathan, T., Padayodi, E., Buiron, N., Racine, D., Zhang, Z., Marceau, D., Rachik, M., 2019. A new nature of microporous architecture with hierarchical porosity and membrane template via high strain rate collision. Materialia 5, 100205. https://doi.org/10.1016/j.mtla.2018.100205

Sapanathan, T., Ibrahim, R., Khoddam, S., Zahiri, S.H., 2015. Shear blanking test of a mechanically bonded aluminum/copper composite using experimental and numerical methods. Mater. Sci. Eng. A 623, 153–164. https://doi.org/https://doi.org/10.1016/j.msea.2014.11.045

Sapanathan, T., Raoelison, R.N., Buiron, N., Rachik, M., 2017. In situ metallic porous structure formation due to ultra high heating and cooling rates during an electromagnetic pulse welding. Scr. Mater. 128, 10–13. https://doi.org/10.1016/j.scriptamat.2016.09.030

Sapanathan, T., Jimenez-Mena, N., Sabirov, I., Monclús, M.A., Molina-Aldareguía, J.M., Xia, P., Zhao, L., Simar, A., 2019. A new physical simulation tool to predict the interface of dissimilar aluminum to steel welds performed by friction melt bonding. J. Mater. Sci. Technol. 35, 2048–2057. https://doi.org/10.1016/j.jmst.2019.05.004

Wang, K., Shang, S.-L., Wang, Y., Vivek, A., Daehn, G., Liu, Z.-K., Li, J., 2020. Unveiling non-equilibrium metallurgical phases in dissimilar Al-Cu joints processed by vaporizing foil actuator welding. Mater. Des. 186, 108306.

https://doi.org/10.1016/J.MATDES.2019.108306

Wang, X, Wang, X. J, Li, F., Lu, J., Liu, H., 2020. Interface Kinematics of Laser Impact Welding of Ni and SS304 Based on Jet Indentation Mechanism. Metall. Mater. Trans. A. https://doi.org/10.1007/s11661-020-05733-0

Watanabe, M., Kumai, S., 2009. Interfacial morphology of magnetic pulse welded aluminum/aluminum and copper/copper lap joints. Mater. Trans. 50, 286–292. https://doi.org/10.2320/matertrans.L-MRA2008843

Zhang, Y., Babu, S.S., Prothe, C., Blakely, M., Kwasegroch, J., LaHa, M., Daehn, G.S., 2011. Application of high velocity impact welding at varied different length scales. J. Mater. Process. Technol. 211, 944–952. https://doi.org/10.1016/J.JMATPROTEC.2010.01.001